\documentclass[preprint2]{aastex701}
\usepackage{mhchem}

\begin{document}

\title{The observation of Ground Level Enhancement GLE 77 by the neutron detectors of the Experimental Complex NEVOD}

\author[orcid=0009-0008-8275-9235,sname='Volkov']{Evgenii Volkov}
\affiliation{National Research Nuclear University MEPhI (Moscow Engineering Physics Institute)}
\email[show]{eugenvolk00@gmail.com}
\correspondingauthor{Evgenii Volkov}

\author[orcid=0009-0002-4713-8800, sname='Chelidze']{Kseniia Chelidze} 
\affiliation{National Research Nuclear University MEPhI (Moscow Engineering Physics Institute)}
\email{KSChelidze@mephi.ru}

\author[orcid=0009-0000-2551-0857, sname='Gromushkin']{Dmitrii Gromushkin} 
\affiliation{National Research Nuclear University MEPhI (Moscow Engineering Physics Institute)}
\email{DMGromushkin@mephi.ru} 

\author[orcid=0000-0002-1317-2851, sname='Khokhlov']{Semen Khokhlov} 
\affiliation{National Research Nuclear University MEPhI (Moscow Engineering Physics Institute)}
\email{SSKhokhlov@mephi.ru}

\author[orcid=0009-0009-1412-2846, sname='Khomchuk']{Evgenii Khomchuk} 
\affiliation{National Research Nuclear University MEPhI (Moscow Engineering Physics Institute)}
\email{EPKhomchuk@mephi.ru} 

\author[orcid=0000-0001-6820-0250, sname='Petrukhin']{Anatoly Petrukhin} 
\affiliation{National Research Nuclear University MEPhI (Moscow Engineering Physics Institute)}
\email{AAPetrukhin@mephi.ru} 

\author[orcid=0000-0003-1697-5024, sname='Shulzhenko']{Ivan Shulzhenko} 
\affiliation{National Research Nuclear University MEPhI (Moscow Engineering Physics Institute)}
\email{IAShulzhenko@mephi.ru} 

\begin{abstract}

A Ground Level Enhancement event was observed by neutron detectors designed for the registration of extensive air showers at the Experimental Complex NEVOD. The potential for that was unlocked by a recent modernization of the experimental setup that included implementation of additional channels for measuring neutron flux variation. At 10:15 UT on November 11, 2025, a sudden and significant increase in the neutron flux was detected by two installations: PRISMA-36 and URAN arrays. For the first time, a GLE has been recorded using a set of neutron detectors oriented at the extensive air shower studies. We present the measured EC NEVOD data and the results of the preliminary analysis of the observed GLE. 

\end{abstract}


\section{Introduction} \label{sec:intro} 

Ground level enhancements (GLE) are generally recognized as a simultaneous increase in the counting rates reported by several neutron monitors (\cite{Poluianov, Kammeemoon_GLE}). GLE events are also preceded and accompanied by an increase in primary cosmic proton flux recorded by satellites (\cite{Poluianov_SEP}).

Information obtained from a combined analysis of data from different neutron monitors usually is used to reconstruct the characteristics of a primary solar flux, such as the solar energetic particle rigidity spectrum and the angular distribution of cosmic ray flux (\cite{Larsen_Mishev_car}).

The Global Neutron Monitor (NM) Network recorded notable increases with the onsets between 10:10 and 10:20 UT November 11, 2025 based on the data from \url{https://www.nmdb.eu/nest/}. The GLE \# 77 was a result of increased solar activity, producing an X-class flare. 
It was the first event of this type recorded by neutron detectors designed to measure the extensive air showers (EAS) at the Experimental Complex (EC) NEVOD. In this paper, we report the results of neutron flux enhancement measurement by the EC NEVOD neutron detectors.

\section{Experimental setups} \label{sec:exp}

The Experimental Complex NEVOD (\cite{Yashin_2021}) is located on the campus of MEPhI ($55.7^\circ$ N, $37.7^\circ$ E) in Moscow, Russia. At the EC NEVOD two installations for studying neutrons in extensive air showers are being operated: the PRISMA-36 and URAN arrays. The total number of neutron detectors in the EC NEVOD is 108. Their total effective detection area is $39 \text{ m}^2$. These detectors have high sensitivity to thermal neutrons, due to using scintillators $\ce{ZnS(Ag)}$ with addition of neutron capturers $\ce{(^6Li \text{or} ^10B)}$ (\cite{Stenkin}). Accordingly, the facilities allow for the study of neutron flux variations.  

The PRISMA-36 array (\cite{AMELCHAKOV_PRISMA}) is designed for EAS neutron studies and located in the experimental hall on the fourth floor of the EC building (173 m a.s.l.), with some detectors installed directly above the water pool of the Cherenkov water deector NEVOD. The effective detection area is about $13 \text{ m}^2$. The absorber above it is made of reinforced concrete with the thickness of $d_{\text{Pr-36}} = 35 ~\text{g/cm}^2$. Each of the 36 detectors of PRISMA-36 is a light-isolated cylindrical polyethylene tank with a $\ce{ZnS(Ag) + ^6LiF}$ scintillator placed at the bottom. To improve light collection by the photomultiplier, a cone with a diffusely reflective coating is located between the detection unit and the scintillator. 

The URAN array (\cite{Gromushkin_2017}), also intended for the EAS research, consists of 72 detectors located on the roofs of the EC NEVOD buildings (180 m a.s.l.). The effective detection area is about $26 \text{ m}^2$. There is no absorber above the detectors. A detector of the URAN array  has the same design as the one for PRISMA-36 but with a $\ce{ZnS(Ag) + B_2O_3}$ scintillator.

The type of detectors used in the PRISMA-36 and URAN arrays have already proved their efficiency in recording and studying EAS neutrons (\cite{Gromushkin_2014, Bingbing_ENDA, Shchegolev}). 

As a result of the 2024 modernization, the additional channels were added for studying neutron flux variations, implemented by recording the signal from the photomultiplier anode to a separate electronics unit.

\section{Data processing and analysis} \label{sec:res}
Registration systems of both facilities have identical electronics. They are synchronized with each other with the accuracy of 10 ns and referenced to global time. Detectors operate independently of each other and record oscillograms to a database for further processing and selection of pulses, when the detection threshold is exceeded on any of the detectors.

A distinctive feature of an engaged scintillator is the difference in decay times as it is detecting charged particles and neutrons. This characteristic allows us to use pulse-shape discrimination to select neutron signals.

Neutron detectors of the EC NEVOD use an analog integration method, so we can obtain integrated neutron pulses with a longer rise time and duration, if compared to pulses from noise signals or charged particles. A detailed description of the data processing and preparation procedures is provided in ref. (\cite{AMELCHAKOV_PRISMA}).

\section{Results} \label{sec:res}
\begin{figure}[ht!]
\begin{center}
    \includegraphics[width=1\linewidth]{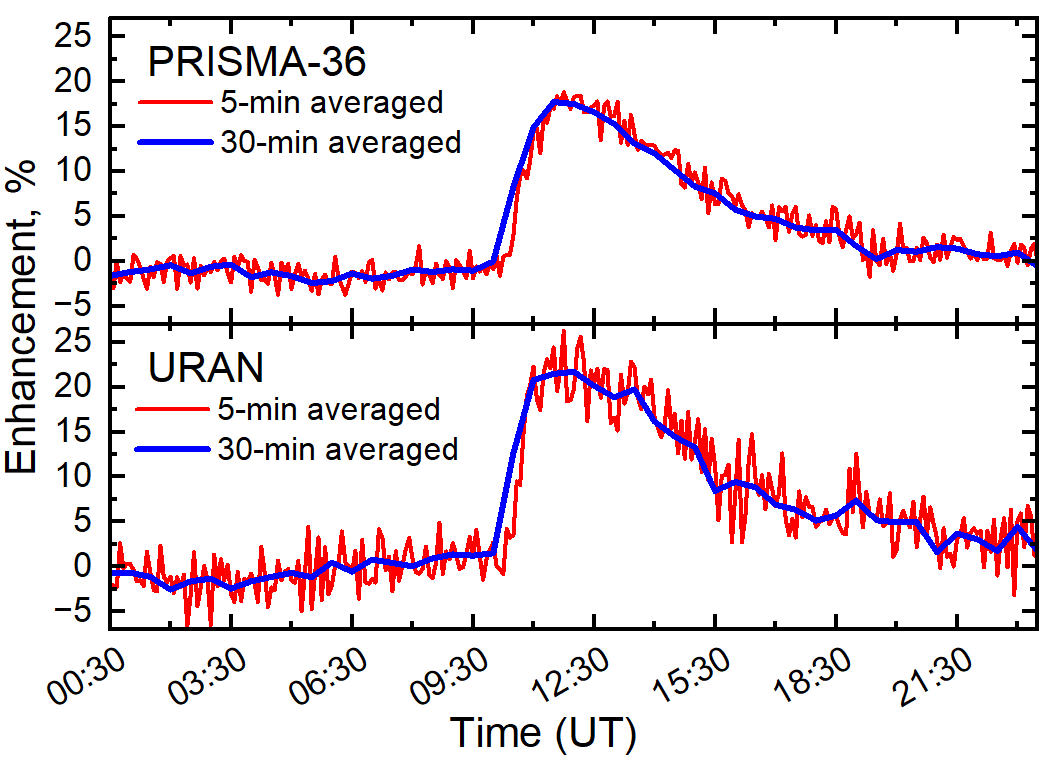}
    \caption{Time profiles of a neutron flux enhancement observed by neutron detectors of the EC NEVOD on November 11, 2025. Data for PRISMA-36 and URAN are shown with averaging done over 5- and 30-minute intervals. 
\label{fig:all_gle}} 
\end{center}
\end{figure}

The results of the neutron flux variations measurements by the EC NEVOD neutron detectors during GLE 77 are demonstrated in Figure \ref{fig:all_gle}. The PRISMA-36 and URAN data are averaged over all detectors of the arrays. 

On November 11, 2025, the outset of GLE was observed on all detectors almost simultaneously at about 10:15 UT (see Figure \ref{fig:all_gle}). On PRISMA-36, the relative increase in the counting rate reached a maximum of $20.1\% \pm 1.1$\%, while URAN exhibited a slightly higher peak enhancement of $24.8\% \pm 1.3$\%. All values were derived from 5-minute averaged data. The times of the peak intensities on both detectors coincided within the measurement uncertainties, with the maximum being reached at 11:45 UT.

To calculate the amplitude of the counting rate increase, 5-minute averaged data for all the detectors were used. To estimate the experimental uncertainties, a statistical error for background neutron counting rate was taken.

\begin{figure}[ht!]
    \plotone{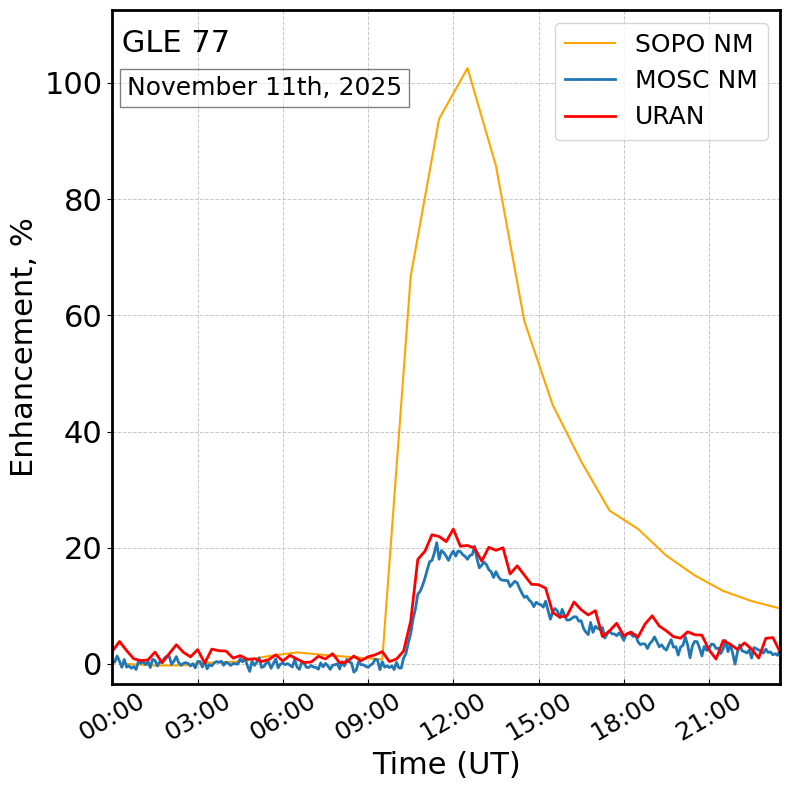}
    \caption{GLE \# 77 based on the data from the SOPO NM, MOSC NM, and URAN on November 11, 2025.
    \label{fig:GLE60}}
\end{figure}

\section{Discussion} \label{sec:disc}

The GLE 77 event was produced by an intense X5.1 solar flare recorded by the GOES X-ray satellite \url{https://www.spaceweatherlive.com/en/solar-activity/solar-flares.html}. The flare started at 09:49 UT and reached its maximum at 10:04 UT on November 15, 2025. As a result, a GLE could be observed in the neutron monitor network starting from 10:10 UT. The results obtained from our detectors can be directly compared with the measurements from the neutron monitor located approximately 30 km away, which operates under nearly identical geomagnetic and atmospheric conditions. The Moscow neutron monitor registered an increase of about $20.9\% \pm 0.5$\%. One of the largest increases was recorded at the SOPO NM, in which the enhancement reached approximately $102.5\% \pm 0.1$\%. Based on the profile of the enhancement with maximal increase (Figure \ref{fig:GLE60}) and according to the classification presented in \citet{Chelidze}, the event is categorized as a Classic (Narrow) GLE.

Comparing the data from the EC NEVOD neutron detectors and the MOSC NM station, the GLE signatures clearly coincide in shape, amplitude, and duration. We consider it as strong evidence that the neutron detection systems of the EC NEVOD can effectively register events such as GLEs.

Similarly, in our previous article on Forbush decrease recordings, we demonstrated an excellent agreement of the results received from the EC NEVOD with the results from the neutron monitors (\cite{Volkov}).

Foregoing indicates the possibility of using the neutron detectors of the EC NEVOD for monitoring the neutron flux variations from cosmic rays with an accuracy comparable to that of neutron monitors due to the large number of detectors and the effective recording area, as well as, the potential inclusion of our detectors into the Global Neutron Monitor Network.

\section{Acknowledgments}
This work was performed at the Experimental Complex NEVOD within the frameworks of the Project FSWU-2026-0011.

\bibliography{biblio.bib}{}
\bibliographystyle{aasjournalv7}

\end{document}